\documentclass[
    ,final            
  ]
  {aipproc}

\layoutstyle{6x9}

\begin{document}

\title{Lattice QCD meets experiment in hadron physics}

\author{Christine Davies}{
  address={Dept. of Physics and Astronomy, University of Glasgow, 
Glasgow, G12 8QQ, UK}
}

\author{Peter Lepage}{
  address={Laboratory for Elementary-Particle Physics, Cornell University,Ithaca, NY 14853, USA}
}

\begin{abstract}
We review recent results in lattice QCD from numerical simulations that allow
for a much more realistic QCD vacuum than has been possible before. Comparison 
with experiment for a variety of hadronic quantities gives agreement to within 
statistical and systematic errors of 3\%. We discuss the implications of this 
for future calculations in lattice QCD, particularly those which will provide 
input for $B$ factory experiments. 
\end{abstract}

\maketitle


\section{Introduction}

QCD is a key component of the Standard Model of particle physics. 
It gives us a rich spectrum of bound states of quarks and gluons
whose properties are predictable from QCD if we can solve the theory. 
QCD is strongly coupled in this regime, however, and we need the 
non-perturbative techniques of lattice QCD to do this from first 
principles. 

Most (but not all) questions which lattice QCD can address require 
calculations with a precision of a few percent to answer them. 
These include the spectrum of hadrons, their internal 
structure and decay rates. 
In particular, the hunt for internal inconsistencies in the Standard 
Model which could lead to new physics requires calculations 
of hadronic weak matrix elements to 2-3\% to match the 
experimental errors that will become possible. 

\begin{figure}
  \includegraphics[height=.3\textheight]{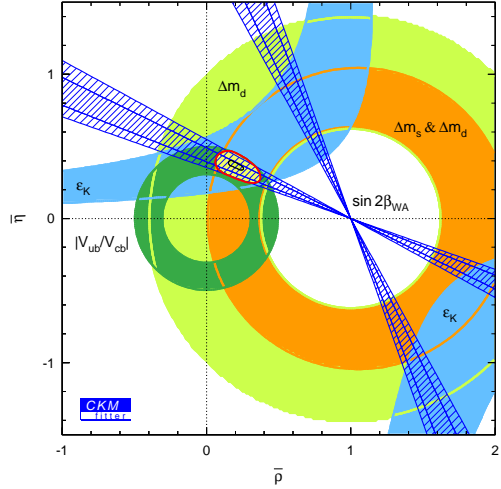}
  \caption{Recent constraints on the unitarity triangle 
from CKMfitter~\cite{ckmfitter}}
\label{figuni}
\end{figure}

Figure~\ref{figuni} shows the recent status of the combined 
experimental and theoretical efforts to pin down the 
vertex of the Cabibbo-Kobayashi-Maskawa unitarity 
triangle~\cite{ckmfitter}. The different circular constraints 
on the vertex come from different decay rates of $B$ and 
$K$ mesons and are found by dividing experimental rates 
by theoretical results obtained in lattice QCD. The 
current constraints are strongly limited by 
current lattice QCD errors of around 20\%. 

Lattice QCD is hard and numerically very expensive. Recent 
progress~\cite{us} has at last made precision calculations look 
possible and we will concentrate on that work and its 
implications in this review. 

\section{Lattice QCD calculations}

Lattice QCD calculations proceed by the discretisation of 
a 4-d box of space-time into a lattice. The QCD Lagrangian is
then discretised onto that lattice. 
The spacing between the points of the lattice, $a$, is $\approx$ 0.1fm 
in current calculations and the length of a side of the box
is $L \approx$ 3.0fm. Thus our simulations can cover energy scales 
from $\approx$ 2 GeV down to $\approx$ 100 MeV.  

The Feyman Path Integral is 
evaluated numerically in a two-stage process.  
In the first stage sets of gluon fields (`configurations') are 
created which are representative `vacuum snapshots'. In the second stage, 
quarks are allowed to propagate on these background gluon field
and hadron correlators are calculated. The dependence of 
the correlators on lattice time is exponential.
From the exponent
the masses of hadrons of a particular $J^{PC}$ can be 
extracted, and from the amplitude, simple matrix elements.

QCD as a theory has a number of unknown parameters, 
the overall dimensionful 
scale of QCD ($\equiv$ the bare coupling constant) 
and the bare quark masses. To make predictions, these parameters 
must be fixed from experiment. In lattice QCD we do this by using 
one hadron mass for each parameter. The quantity which is equivalent to 
the overall scale of QCD on the lattice is the lattice spacing. 

Lattice calculations are hard and time-consuming. Progress has 
occurred in the last thirty years through gains in computer power but 
also, often more importantly, through gains in calculational
efficiency and physical understanding. One particular area which 
revolutionised the field from the mid-1980s was the understanding 
of the origin of discretisation errors and their removal by 
improving the lattice QCD Lagrangian. Discretisation errors appear
whenever equations are discretised and solved numerically. They 
manifest themselves as a dependence of the physical result on 
the unphysical lattice spacing. In lattice QCD, as elsewhere, 
they are corrected by the adoption 
of a higher order discretisation scheme. The complication in 
a quantum field theory like QCD is the presence of radiative
corrections to the coefficients in the higher order scheme which 
must be determined (using perturbation theory, for example~\cite{howard}). 

Physical understanding of heavy quark physics on the lattice has 
also made a huge difference to the feasibility of calculating 
matrix elements relevant to the $B$ factory programme on the 
lattice. The use of non-relativistic effective theories requires the 
lattice to handle only scales appropriate to the physics of the 
non-relativistic bound states and not the (large) scale associated 
with the $b$ quark mass. 
$B$
physics is now one of the areas where lattice QCD can make most 
impact.   

One area which has remained problematic, but which this year's results
have addressed successfully, is the handling of light quarks on 
the lattice. In particular the problem is that of how to include 
the dynamical (sea) $u/d/s$ quark pairs that appear as a result of 
energy fluctuations in the vacuum. We can safely ignore $b/c/t$ quarks in the 
vacuum because they are so heavy, 
but we know that light quark pairs have significant effects, for example
in screening the running of the coupling constant and in generating Zweig-allowed
decay modes for unstable mesons. 

Because quarks are fermions, they cannot be simulated directly on 
the computer, but must be `integrated out' of the Feynman Path 
Integral. This leaves an QCD Lagrangian in terms 
of gluon fields which includes ln(det($M$)) where $M$ is 
an enormous ($10^7 \times 10^7$) sparse matrix.  
The inclusion of dynamical quarks is then numerically 
very expensive, particularly 
as the quark mass is reduced towards the small values which we know 
the $u$ and $d$ quarks have. 

Many calculations even today use the `quenched approximation' in which 
the light quark pairs are ignored. Results then suffer from a 
systematic error of $\cal{O}$(20\%). A serious problem with the quenched 
approximation is the lack of internal consistency which means that the 
results depend on the hadrons that were used to fix the parameters 
of QCD. 
This ambiguity plagues the lattice 
literature. 

Other calculations have included 2 flavours of degenerate dynamical 
quarks, i.e. $u$ and $d$, but with masses 10-20$\times$ the 
physical ones. This approximation is better than the 
quenched approximation but large uncertainties remain because 
the $s$ quark is omitted.
Results must also be extrapolated to the physical $u/d$ quark mass and 
chiral perturbation theory is a good tool for this. However, chiral 
perturbation theory only works well if the $u/d$ quark mass is light 
enough and, for errors at the few percent level, this means less 
than $m_s/2$. This has been impossible to achieve in most calculations.   

New results this year~\cite{us} have included $u, d$ and $s$ quarks in 
the vacuum, with light enough $u/d$ masses to perform accurate 
chiral extrapolations. The results use a new discretisation 
of the quark action - the numerically fast 
improved staggered formalism. This formalism is well-matched 
to the supercomputing power of a few Tflops that is currently 
achievable.  

\subsection{Improved staggered quarks}

The starting point for the staggered quark formalism is the naive
discretisation of the Dirac quark action onto a lattice. This 
action has good features: chiral symmetry and discretisation 
errors that appear only as the square and higher powers of the lattice 
spacing. The naive discretisation suffers from the notorious 
doubling problem, however. A single quark species on the lattice 
gives rise to $16$ quark species, or tastes, on a 4-d lattice. The 
additional tastes appear around the edges of the Brillioun zone, 
where $p \approx \pi/a$, as copies of a $p \approx 0$ quark. 
This would not be a problem if there were no interaction between 
the different tastes since the quark action would then fall 
apart into 16 different pieces in an appropriate basis 
and we could take ${\rm det}(M)^{(1/16)}$
in simulations to give the effect of 1 quark flavour. 

There is interaction between the different tastes, however. It 
is mediated by highly virtual gluons, with momenta around $\pi/a$.
A quark of one taste can absorb or emit such a high 
momentum gluon and turn into a quark of another taste. 
The effects of this taste-changing interaction are 
quite severe for the naive action, giving rise to large 
discretisation errors (even though formally of $\cal{O}$$(a^2)$)
and large perturbative renormalisation factors, e.g. for 
the quark mass, when translating from the 
lattice scheme to the continuum. The degeneracy in 
mass of mesons made from quarks of different taste 
is lost. This is most noticeable for the pions because 
there is a light Goldstone boson.

Because the taste-changing interaction is a high momentum one it 
can be understood in lattice perturbation theory.
In particular, the effects can be significantly improved 
by suppressing the coupling of quarks to gluons 
of momenta $\pi/a$ in any direction. This is 
achieved by `smearing' the gluon field in the action 
in a particular way~\cite{gpl, orginos}, and can be thought of as part 
of the standard Symanzik programme for systematically 
removing discretisation errors from lattice actions.  

It is simple to `stagger' the naive action and its improved 
variant to remove an exact 
degeneracy of a factor of 4 in tastes which arises from 
the spin degree of freedom. This results in an action 
with 4 doublers which can be simulated on the 
lattice using ${\rm det}(M)^{(1/4)}$ per flavour. It is very 
fast numerically because there is only one spin 
degree of freedom per site and the eigenvalues of $M$ 
are well behaved. 
This is what has allowed the MILC collaboration to 
generate ensembles of configurations which include 
$u$, $d$, and $s$ quarks in the vacuum with 
much more realistic masses than before~\cite{milc}. 

Some worries remain about potential non-locality in 
the action as the result of taking the fourth root. 
However, this causes no problem in perturbative 
QCD where a simple power series in $x$ is obtained 
for an action with ${\rm det}(M)^x$. Stringent non-perturbative 
tests are also then needed. Luckily these tests are possible 
in this formalism with present day computers because of 
its speed, and are exactly the calculations required 
to test (lattice) QCD. The results, shown in the 
next section, speak for themselves. 

\section{Recent results}

The MILC collaboration have made sets of ensembles of 
gluon field configurations which include 2 degenerate
light dynamical quarks ($u,d$) and 1 heavier 
one ($s$)~\cite{milc}. Taking the $u$ and $d$ masses the 
same makes the lattice calculation much faster and 
leads to negligible errors in isospin-averaged 
quantities. The dynamical $s$ quark mass is chosen 
to be approximately correct based on earlier studies 
(in fact the subsequent analysis shows that it was slightly 
high and further ensembles are now being made with a lower value). 
The dynamical 
$u$ and $d$ quarks take a range of masses down 
as low as a sixth of the (real) $m_s$.
The sets of ensembles divide into two different 
values of the lattice spacing, 0.13fm and 0.09fm, 
and the spatial lattice volume is $(2.5{\rm fm})^3$,
reasonably large.  Analysis of hadronic quantities on 
these ensembles has been done by the 
MILC and HPQCD collaborations~\cite{us}. 

There are 5 bare parameters of QCD relevant to this 
analysis: $\alpha_s, m_{u/d}, m_s, m_c$ and $m_b$. 
The lattice spacing takes the place of $\alpha_s$ 
in lattice QCD. 
It is important that these 
parameters are fixed using the masses of `gold-plated' hadrons,
i.e. hadrons which are well below their strong decay thresholds.
Such hadrons are well-defined experimentally and theoretically 
and should be accurately calculable in lattice QCD. 
Using them to fix parameters will not then introduce 
unnecessary additional 
systematic errors into lattice results for other quantities. 
This has not always been done in past lattice 
calculations, particularly in the quenched approximation.
It becomes an important issue when lattice QCD is 
to be used as a precision calculational tool. 
We use the radial excitation energy in the $\Upsilon$ 
system (i.e. the mass splitting between the 
$\Upsilon^{\prime}$ and the $\Upsilon$) to fix the 
lattice spacing and $m_{\pi}$, $m_K$, 
$m_{D_s}$ and $m_{\Upsilon}$ to fix the quark masses. 

\begin{figure}
  \includegraphics[height=.3\textheight]{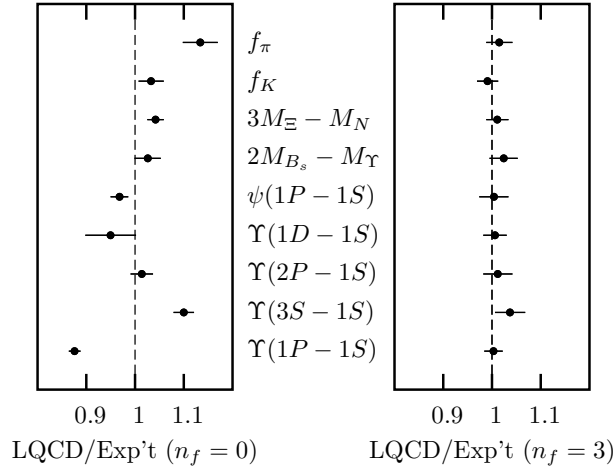}
  \caption{Lattice QCD results divided by experiment for 
a range of `gold-plated' quantities which cover the full 
range of hadronic physics ~\cite{us}. The unquenched 
calculations on the right show agreement with experiment 
across the board, whereas the quenched approximation 
on the left give systematic errors of $\cal{O}$(10-20\%). }
\label{ratio}
\end{figure}

We can then focus on the calculation of other 
gold-plated masses and decay constants. 
If QCD is correct and lattice QCD is to work it must reproduce the 
experimental results for these quantities precisely. 
Figure~\ref{ratio} shows that this indeed works 
for the unquenched calculations with $u, d$ and $s$
quarks in the vacuum. A range of gold-plated hadrons 
are chosen which range from decay constants for light hadrons through 
heavy-light masses to heavyonium. This tests QCD in 
different regimes in which the sources of systematic 
error are very different and stresses the point that 
QCD predicts a huge range of physics with a small 
set of parameters.

\begin{figure}
\includegraphics[height=.32\textheight]{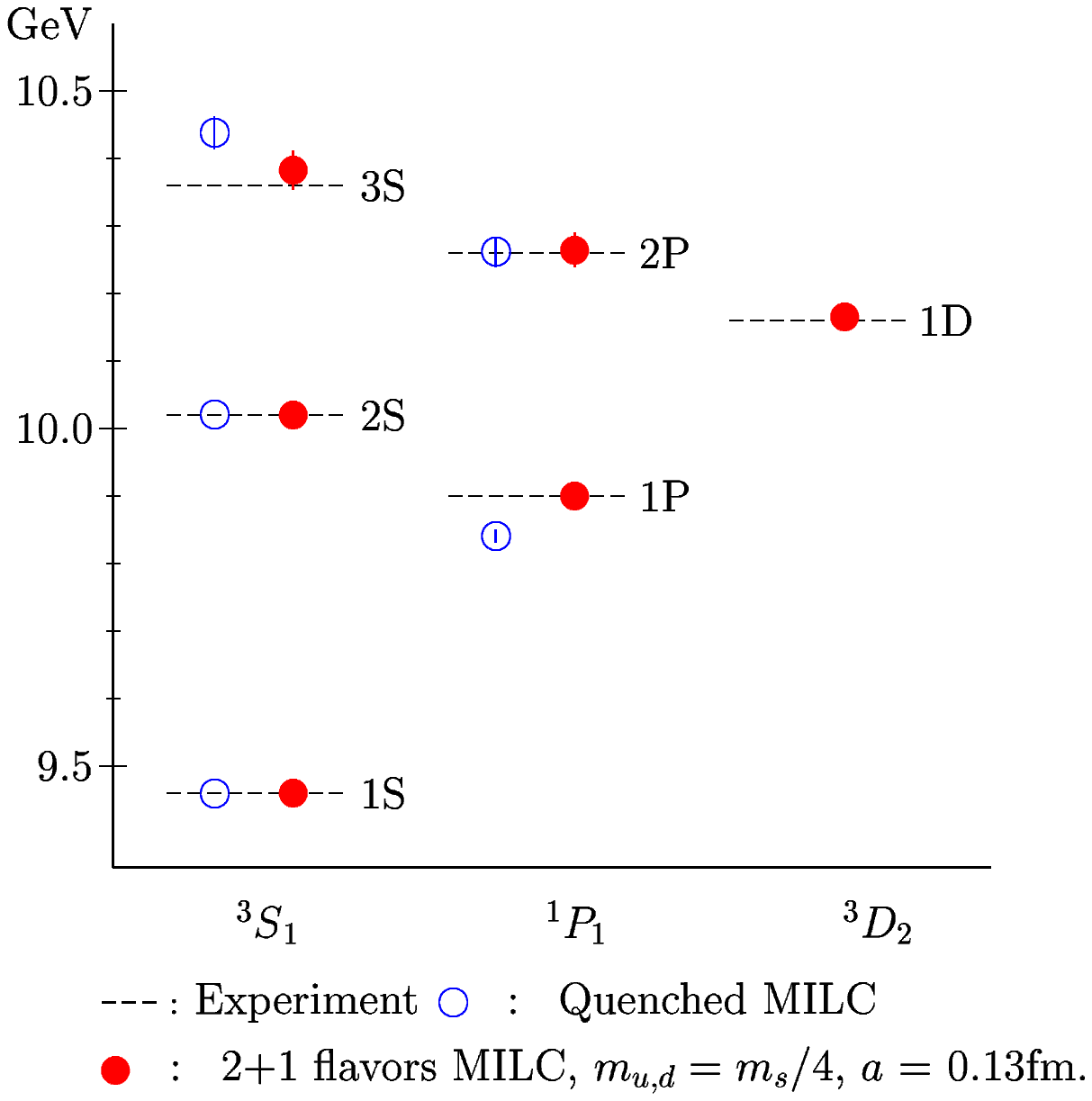}
\includegraphics[height=.32\textheight]{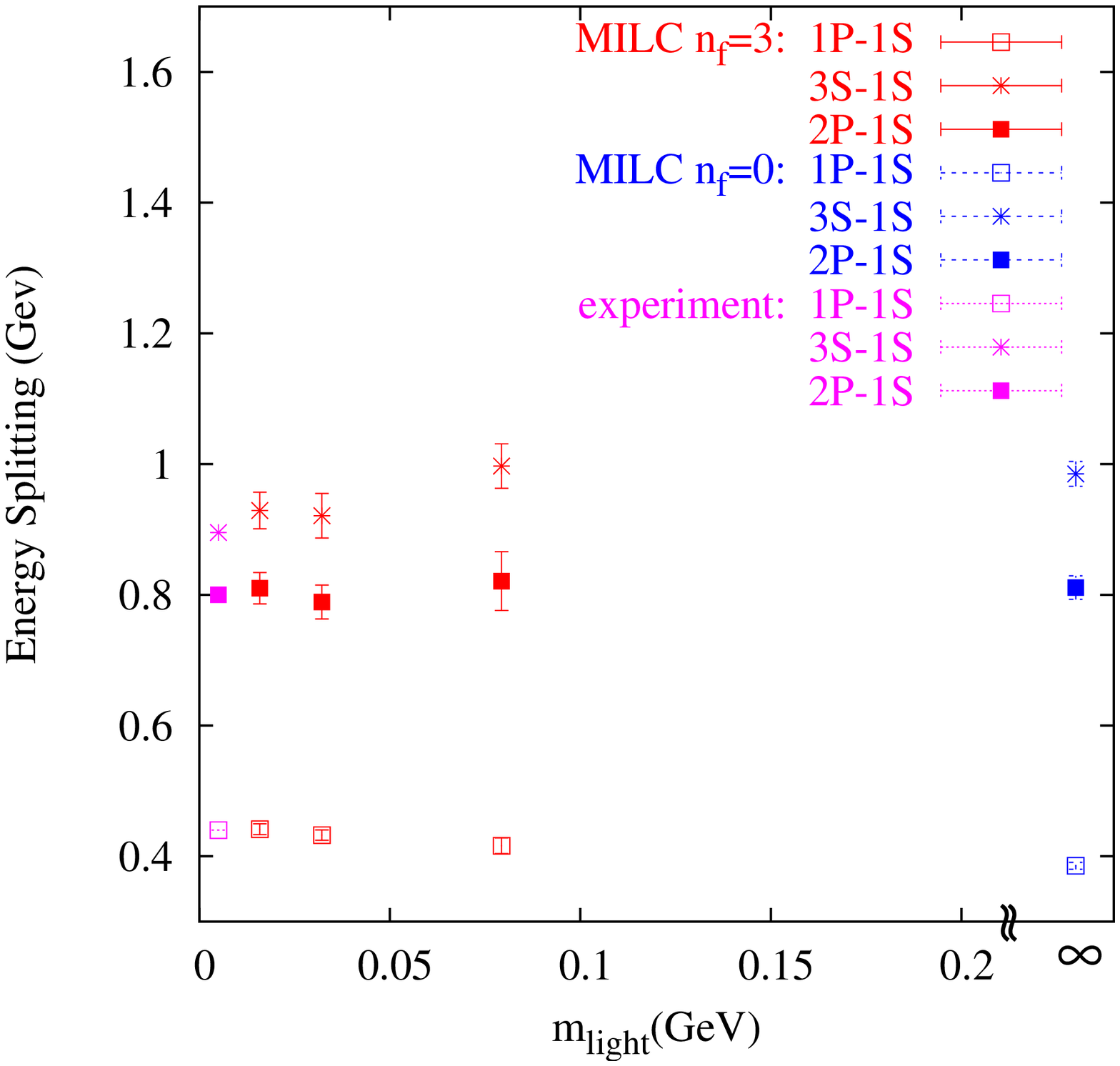}
  \caption{Radial and orbital splittings in the $\Upsilon$ system 
from lattice QCD in the quenched approximation and including 
$u, d$ and $s$ dynamical quarks. In this plot the lattice spacing 
was fixed from the radial excitation energy i.e. the splitting 
between the $\Upsilon^{\prime}$ and the $\Upsilon$ and
the $b$ quark mass was tuned to get the $\Upsilon$ mass correct. 
The right hand plot shows these splittings plotted as a function 
of the bare dynamical $u/d$ quark mass for several ensembles 
of MILC configurations. The leftmost lattice points are the ones used 
in the left-hand plot and Figure~\ref{ratio}. } 
\label{ups}
\end{figure}

References~\cite{gottlieb, gray, charm, bernardlat03} 
give more details on the 
quantities shown in Figure~\ref{ratio}. Here we will discuss 
some of these. Figure~\ref{ups} shows 
the radial and orbital splittings in the $b\overline{b}$ 
($\Upsilon$) system for the quenched approximation ($n_f$ = 0)
and with the dynamical MILC configurations with  
3 flavours of dynamical quarks. 
Our physical understanding of the $\Upsilon$ system is very good 
and there are a lot of gold-plated states well below 
decay thresholds, which makes it a valuable system for lattice QCD 
tests. We use the standard lattice NRQCD effective theory for the valence 
$b$ quarks, which takes advantage of the non-relativistic nature 
of the bound states. The lattice NRQCD action is accurate through 
$v^4$ where $v$ is the velocity of the $b$ quark in its bound state. 
This means that spin-independent splittings, such as radial and 
orbital excitations, are simulated through next-to-leading-order 
and should be accurate to $\approx$ 1\%. Thus the test of 
QCD using these splittings is a very accurate one. 
The fine structure in the spectrum is only correct through leading-order 
at present and more work must be done to bring this to the 
same level and allow tests against, for example, the splittings 
between the different $\chi_b$ states~\cite{gray}. 

The $\Upsilon$ system is a good one for looking at the effects 
of dynamical quarks because we do not expect it to be very 
sensitive to dynamical quark masses. The momentum transfer 
inside an $\Upsilon$ is larger than any of the $u,d$ or $s$ masses 
and so we expect the radial and orbital splittings to simply 
`count' the number of dynamical quarks once we have reasonably 
light dynamical quark masses. The righthand plot of 
Figure~\ref{ups} shows this to 
be true - the splittings are independent of the dynamical 
$u/d$ quark mass in the region we are working in (and 
therefore for the points plotted in the left hand figure of 
Figure~\ref{ups} and in Figure~\ref{ratio}).  

\begin{figure}
  \includegraphics[height=.3\textheight]{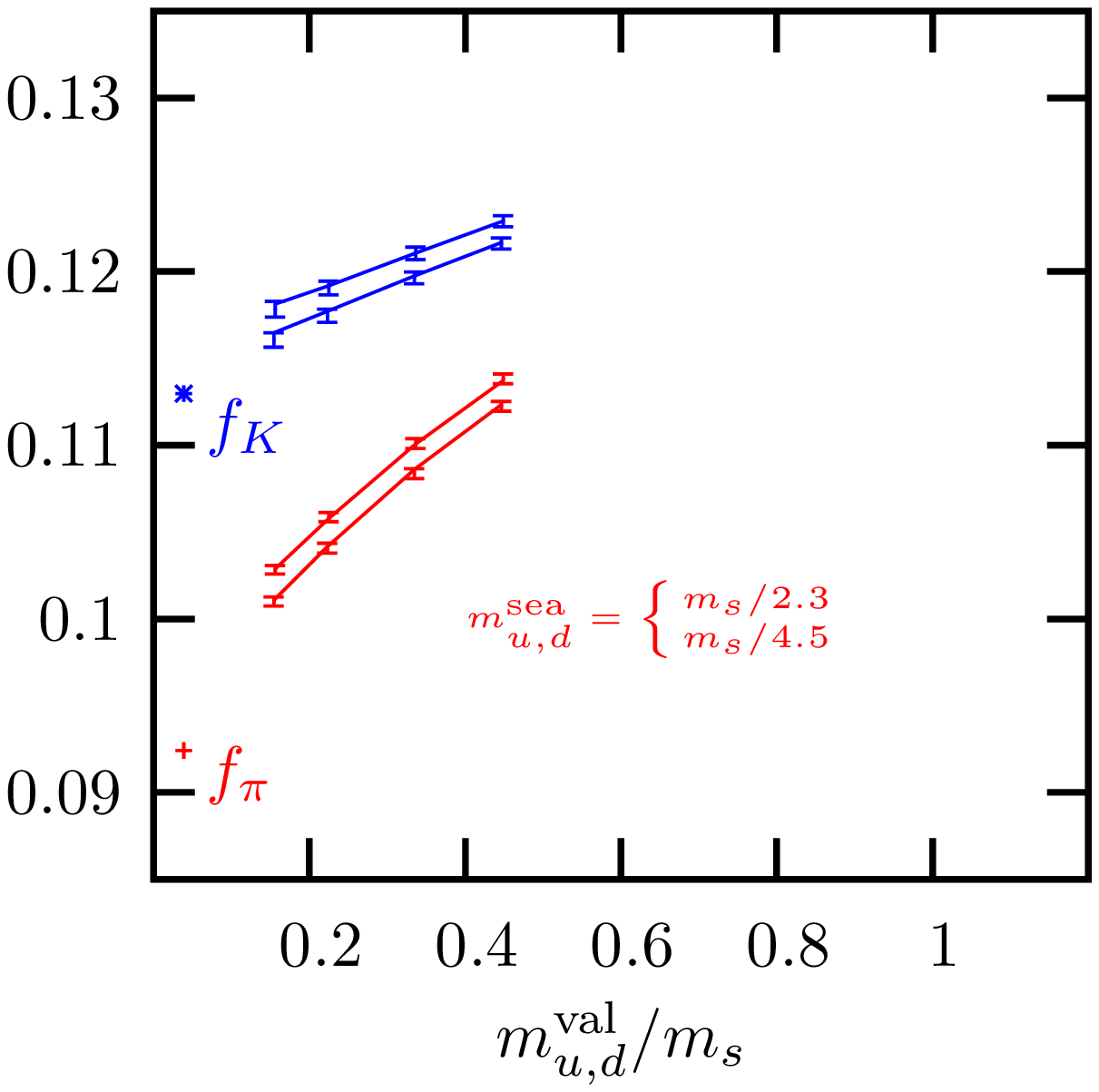}
  \includegraphics[height=.3\textheight]{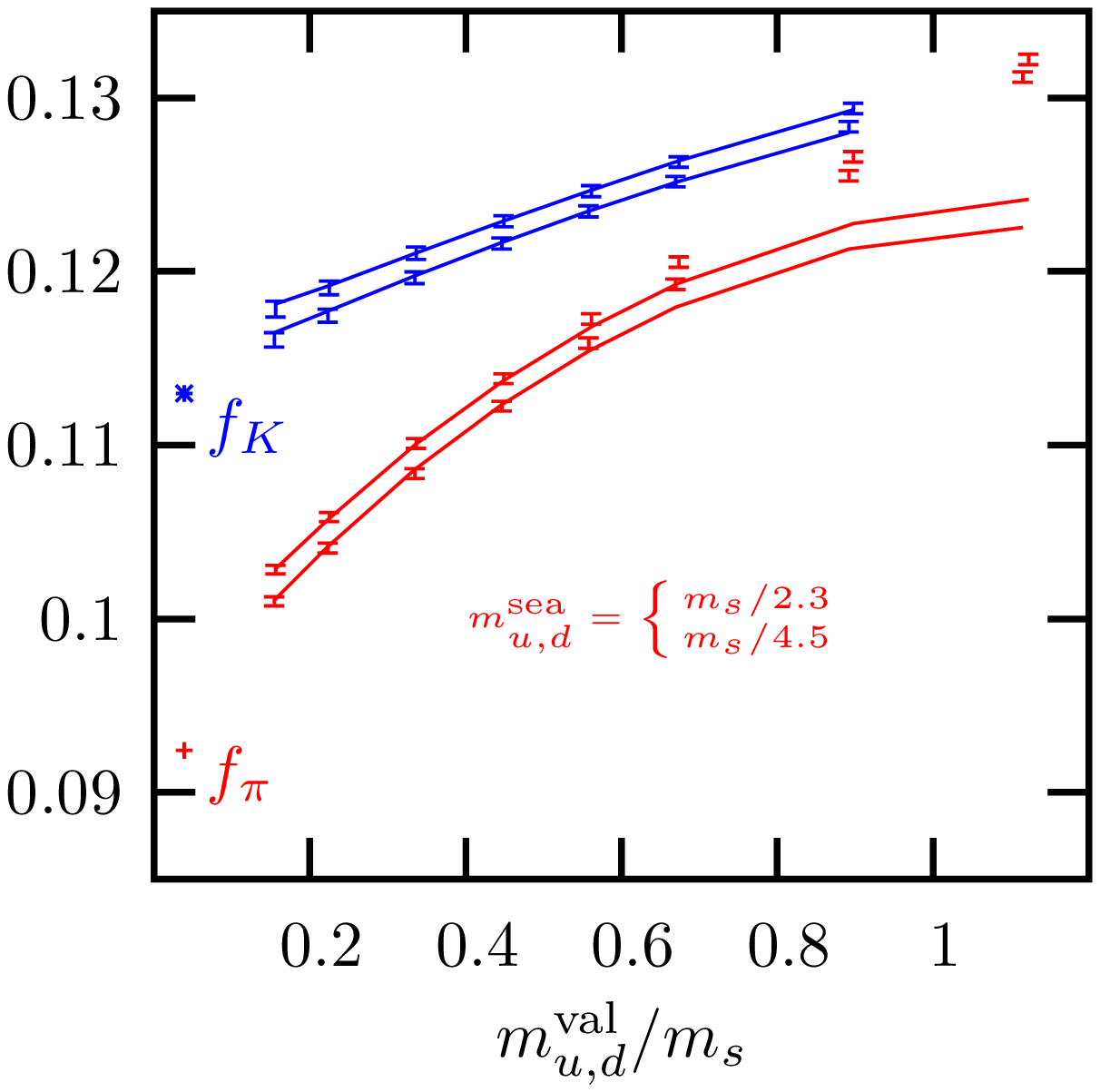}
  \caption{Results for the $\pi$ and $K$ decay constants 
as a function of light quark mass for two dynamical MILC ensembles 
at a lattice spacing of 0.09fm. The plot on the left shows 
the chiral extrapolation using only results with valence $u/d$ quark 
masses $< m_s/2$~\cite{us}. The chiral extrapolation must subsequently 
be corrected for the incorrect valence and sea $s$ quark mass 
to give the results in Figure~\ref{ratio}. 
The plot on the right shows that this chiral fit
from light $u/d$ quark masses does not agree well with the 
data for $m_{u/d} > m_s/2$.
}
\label{fpifk}
\end{figure}

The $\pi$ and $K$ decay constants are important 
light hadron matrix elements, related to the purely 
leptonic decay rate via a $W$, and experimentally 
well-known. These are very sensitive to light 
quark masses and require a well-controlled 
extrapolation in the $u/d$ quark mass and interpolation 
in the $s$ quark mass to get accurate results to 
compare to experiment. Chiral perturbation theory 
can be used to perform the $u/d$ quark mass 
extrapolation provided the masses used on the 
lattice are small enough for the expansion in 
powers of quark mass ($\equiv m_{\pi}^2/(1{\rm GeV}^2)$) 
and its logarithms to 
work well. In practise this means that second 
order chiral perturbation theory should work 
at the 2\% level for $m_{u/d} < m_s/2$.  
Note that the error is set by the largest quark mass 
used in the chiral fits, {\it not} the smallest. 

Figure~\ref{fpifk} shows the results and chiral extrapolation 
for the decay constants on the ensembles of MILC configurations 
with $m_{u/d}^{sea} = m_s/2.3$ and $m_s/4.5$ at a lattice 
spacing of 0.09 fm. The curves in the left plot show the 
chiral extrapolation using only results with $m_{u/d}^{valence} < m_s/2$.  
This extrapolation has to be corrected, using the lattice 
results, to interpolate to the physical $s$
quark mass for both sea and valence $s$ quarks. 
This then gives the results shown in Figure~\ref{ratio} 
which agree with experiment. The plot on the right shows 
what happens when the chiral extrapolation fit obtained 
in the left plot is evaluated for larger valence $m_{u/d}$.
The $f_{\pi}$ results start to show clear disagreement 
for $m_{u/d} > m_s/2$, which makes the problem 
of performing accurate chiral extrapolations using 
results with $m_{u/d} > m_s/2$ obvious. 
Previous lattice calculations have been forced by 
computing cost to work only in 
this regime, with the added problem that the 
sea $m_{u/d}$ is also large~\cite{aoki}. 

\begin{figure}
  \includegraphics[height=.35\textheight]{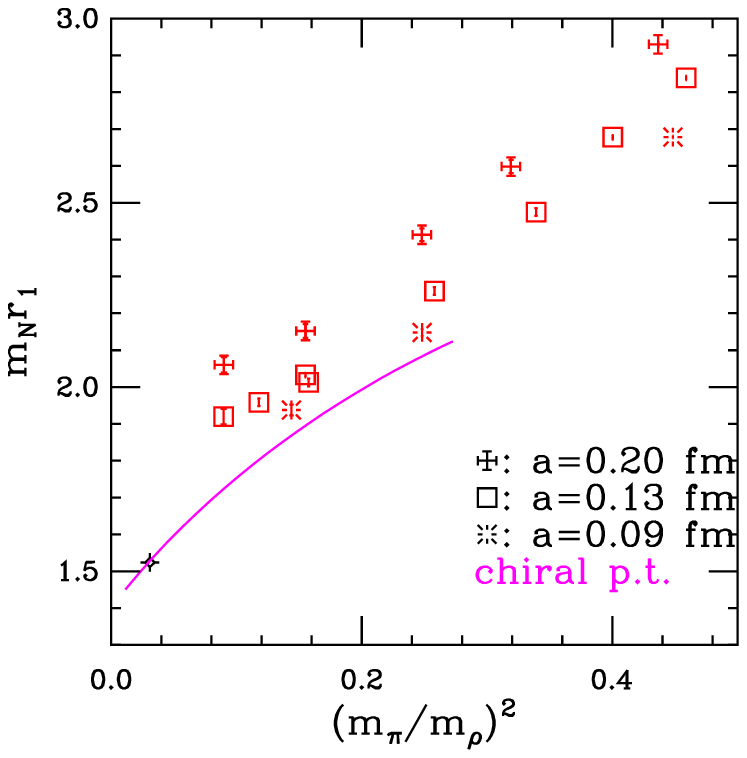}
  \includegraphics[height=.35\textheight]{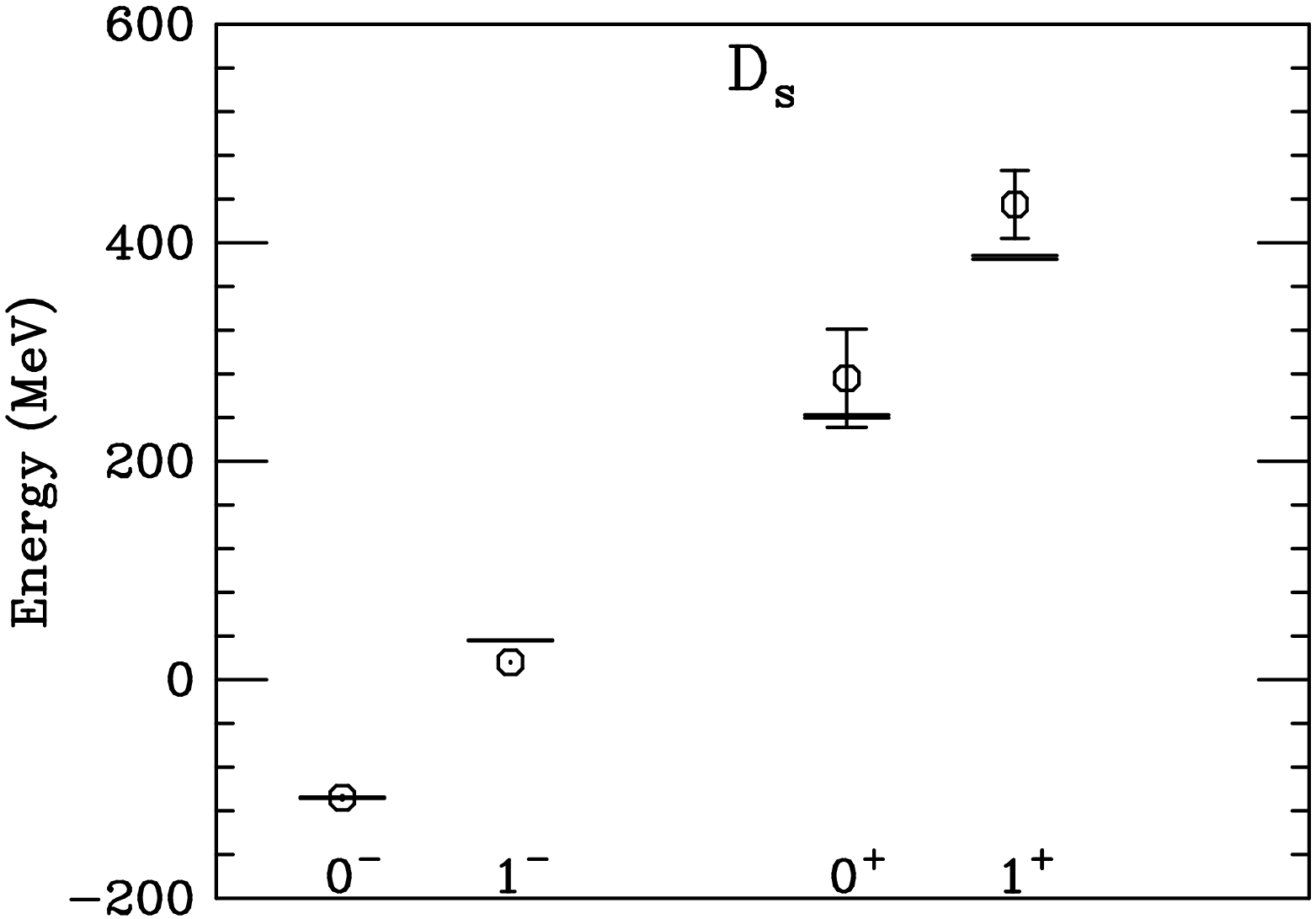}
  \caption{The left-hand plot shows results for the nucleon mass on 
MILC ensembles for different lattice spacings and dynamical 
quark masses. The nucleon mass is given in units of $r_1$, a 
parameter from the heavy quark potential whose physical value 
is 0.32fm. The dynamical quark mass is indicated by the 
variable $m_{\pi}^2/m_{\rho}^2$. The curve roughly indicates 
chiral perturbation theory~\cite{gottlieb}. 
The right-hand plot shows the spectrum of $D_s$ states obtained 
from the MILC dynamical configurations with $m_{u/d} = m_s/4$ 
and lattice spacing 0.13fm~\cite{fnalds}.} 
\label{milcmn}
\end{figure}

Another gold-plated hadron mass is that of the nucleon. A
full chiral extrapolation of results for this on the 
MILC configurations has not yet been done. The left-hand plot 
of Figure~\ref{milcmn}
shows very encouraging signs that an answer in agreement 
with experiment will be found~\cite{gottlieb}. There is a clear sign of 
dependence on the lattice spacing, however, which will have to be 
taken into account. 
Combinations of baryon masses can be made which are 
relatively insensitive to $u/d$ quark masses and 
other effects and 
it is one of these, $3m_{\Xi} - m_N$, which is plotted in Figure~\ref{ratio}. 

It is important to realise that accurate lattice QCD 
results are not going to be obtainable in the near future for every 
hadronic quantity of interest. What these results show 
is that `gold-plated' quantities should now work. Gold-plated
hadrons are those well below decay threshold for strong decays. 
Unstable hadrons, or even those within 100 MeV or so of Zweig-allowed
decay modes, have a strong coupling to their real or virtual 
decay channel which 
is not correctly simulated on the lattice. The problem is that,
with the lattice volumes being used, the 
allowed non-zero momenta are typically greater than 400 MeV 
and this significantly distorts the decay channel contribution. 
Much larger simulations will be necessary to handle these hadrons. 

Gold-plated hadrons include: $\pi$, $K$, $D$, $D_s$, $J/\psi$, 
$\Upsilon$, $B$, $B_s$, $p$, $n$, $\Lambda$, $\Omega$ etc. 
The following are {\it not} gold-plated: $\rho$, $\phi$, $D^*$, 
$D_{sJ}$, $\Delta$, $N^*$, pentaquarks, glueballs and hybrids
in general. Lattice calculations will not get the masses 
right for non-gold-plated hadrons even when light dynamical 
quarks are included. This does not preclude lattice calculations 
giving useful qualitative results and insight but these points 
should be borne in mind for any quantitative comparison. 

Figure~\ref{milcmn} also shows the spectrum 
of $D_s$ states obtained on the dynamical MILC configurations~\cite{fnalds}. 
The valence $c$
quarks are simulated using an effective theory which, in 
a similar way to the $\Upsilon$ above, should be accurate 
for spin-independent splittings and not quite so accurate 
for fine structure in the spectrum. The hyperfine splitting 
between the $D_s$ and $D_s^*$, for example is currently missing 
a radiative correction to the term in the action proportional 
to the spin coupling to the chromo-magnetic field. This is 
being calculated in lattice perturbation theory~\cite{howard}. Also shown 
are the scalar and axial vector orbital excitations 
compared to the recent experimental results for these 
mesons. The lattice calculation is giving a high result, albeit 
with large statistical errors at present. However, a high 
result is consistent with the fact that these mesons 
are not gold-plated and the lattice calculation does not 
currently include correctly the coupling to their decay modes.    

\begin{figure}
  \includegraphics[height=.35\textheight]{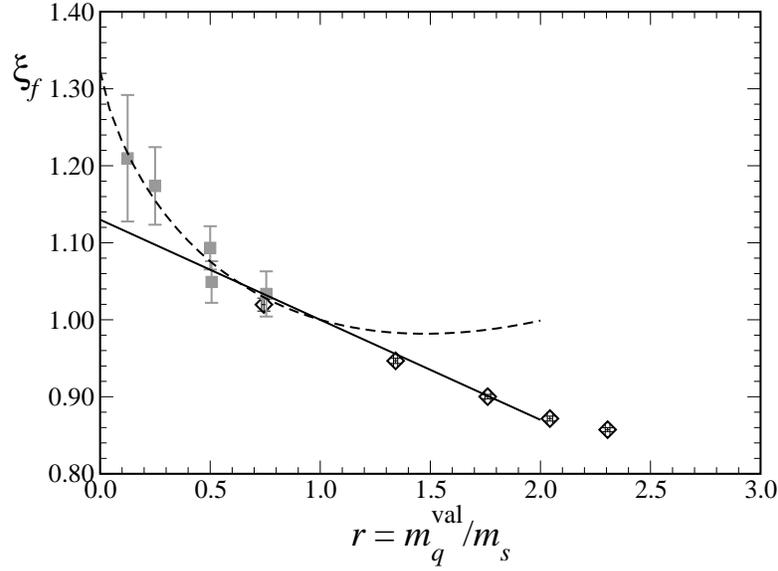}
  \caption{Results for the ratio of $f_{B_s}/f_{B_d}$,
as a function of valence $u/d$ quark mass in units 
of $m_s$~\cite{kronfeld}.  The grey squares are from the dynamical MILC 
ensembles including $u, d$ and $s$ dynamical quarks~\cite{wingatelat03}. 
The black diamonds are from the previous best calculation 
which included 2 flavours of dynamical quarks with 
masses $> m_s2$~\cite{aoki}. The straight line is a 
linear extrapolation for the 2 flavour results, the 
curve includes the possibility of logarithms from 
chiral perturbation theory.  
This ratio, for physical $u/d$ masses, appears in the ratio 
of oscillation frequencies for $B_s$ and $B_d$ mesons, which it is hoped to
measure experimentally.  }
\label{wingate}
\end{figure}

Decay rates which can be accurately calculated for gold-plated 
hadrons are those in which there is at most one 
(gold-plated) hadron in the final state. This therefore 
includes leptonic and semi-leptonic decays and the 
mixing of neutral $B$ and $K$ mesons.  Luckily there is a
gold-plated decay mode available to extract each element 
(except $V_{tb}$)
of the CKM matrix which mixes quark flavours under the 
weak interactions in the Standard Model:
\[ \left( \begin{array}{ccc}
{\bf V_{ud}}  & {\bf V_{us}} & {\bf V_{ub}} \\
\pi \rightarrow l\nu & K \rightarrow l \nu & B \rightarrow \pi l \nu \\
 & K \rightarrow \pi l \nu &  \\
{\bf V_{cd}} & {\bf V_{cs}} & {\bf V_{cb}} \\
D \rightarrow l\nu & D_s \rightarrow l \nu & B \rightarrow D l \nu \\
D \rightarrow \pi l\nu & D \rightarrow K l \nu &  \\
{\bf V_{td}} & {\bf V_{ts}} & {\bf V_{tb}} \\
\langle  B_d | \overline{B}_d \rangle  & \langle B_s | \overline{B}_s \rangle  & \\
\end{array} \right) \]
 As described 
earlier, the determination of the CKM elements and 
tests of the self-consistency of the CKM matrix 
 are the current focus for the search for Beyond the Standard 
Model physics and lattice calculations of these decay rates 
will be a key factor in the precision with which this can be done.  

First calculations on the dynamical MILC configurations have 
concentrated on the $B$ and $B_s$ leptonic decay 
rates~\cite{wingatelat03, fnalds}, because 
these are simplest. They are parameterised by the decay 
constants, $f_B$ and $f_{B_s}$, and these are 
an important 
component of the mixing rate for these mesons, 
which constrains $V_{ts}$ and $V_{td}$. Again one
issue in extracting reliable lattice results for 
$f_B$ and $f_{B_s}$ is the chiral extrapolation in 
the $u/d$ quark mass. Figure~\ref{wingate}
shows results on the MILC configurations for the 
ratio of $f_{B_s}/f_{B_d}$ 
plotted against the valence $u/d$ quark mass~\cite{kronfeld}. 
The data extend into the region $m_{u/d} < m_s/2$ which 
will allow an accurate chiral extrapolation for 
the first time. Although the statistical 
errors are currently rather large, it seems likely that 
the result for this ratio will be larger than previous 
estimates based on extrapolations from larger 
$u/d$ masses, and including only two flavours of dynamical 
quarks~\cite{aoki}. Further calculations of gold-plated matrix 
elements are in progress~\cite{slstuff}. 

\section{Conclusions}

The impact of lattice QCD calculations has been hindered by the 
difficulty of including a realistic QCD vacuum. 
This has led to a level of systematic error far 
greater than the few percent needed to provide input to tests 
of the Standard Model, particularly those testing 
the CKM matrix at $B$ factories. 
New results this year look set to herald a brighter future 
in which accurate calculations, at least for gold-plated 
quantities, are available from the lattice at last.


\begin{theacknowledgments}
We are grateful to PPARC, NSF, DoE and the EU for funding this work,
and to all our collaborators on~\cite{us} for many useful discussions. 
A version of this talk was also presented at the 
2003 Lepton Photon Conference and is published in the 
proceedings of that meeting. 
\end{theacknowledgments}


\bibliographystyle{aipproc}   


\IfFileExists{\jobname.bbl}{}
 {\typeout{}
  \typeout{******************************************}
  \typeout{** Please run "bibtex \jobname" to optain}
  \typeout{** the bibliography and then re-run LaTeX}
  \typeout{** twice to fix the references!}
  \typeout{******************************************}
  \typeout{}
 }

\end{document}